\documentclass[runningheads]{llncs}

\usepackage[ruled,lined,linesnumbered]{algorithm2e}
\usepackage{algorithmic}
\usepackage{amsmath}
\usepackage{amssymb}
\usepackage{array}
\usepackage{blindtext}
\usepackage{bm} 
\usepackage{booktabs} 
\usepackage{caption}
\usepackage{color}
\usepackage{CJKutf8}
\usepackage{enumitem}
\usepackage{graphicx}
\usepackage{hyperref}
\usepackage{multirow}
\usepackage{pifont}
\usepackage[labelformat=simple]{subfig}
\usepackage[labelformat=simple]{subfig}
\usepackage{threeparttable/threeparttable}
\usepackage[normalem]{ulem}
\usepackage{url}

\newcommand{\etal}{\mbox{\emph{et al.}}}

\newcolumntype{P}[1]{>{\centering\arraybackslash}p{#1}}

\begin{document}

\begin{CJK}{UTF8}{gbsn}  

\title{Pixel Embedding Method for Tubular Neurite Segmentation}

\titlerunning{Pixel Embedding Method for Tubular Neurite Segmentation}

\author{
    Huayu Fu \and
    Jiamin Li \and \\
    Haozhi Qu \and Xiaolin Hu \and
    Zengcai Guo
}
\authorrunning{H. Fu, \etal}

\institute{
    Tsinghua University\\
    \email{fuhy22@mails.tsinghua.edu,cn}
}
\maketitle              

\begin{abstract}
Automatic segmentation of neuronal topology is critical for handling large-scale neuroimaging data, as it can greatly accelerate neuron annotation and analysis. However, the intricate morphology of neuronal branches and the occlusions among fibers pose significant challenges for deep learning–based segmentation. To address these issues, we propose an improved framework:

First, we introduce a deep network that outputs pixel-level embedding vectors and design a corresponding loss function, enabling the learned features to effectively distinguish different neuronal connections within occluded regions. Second, building on this model, we develop an end-to-end pipeline that directly maps raw neuronal images to SWC-formatted neuron structure trees. Finally, recognizing that existing evaluation metrics fail to fully capture segmentation accuracy, we propose a novel topological assessment metric to more appropriately quantify the quality of neuron segmentation and reconstruction.

Experiments on our fMOST imaging dataset demonstrate that, compared to several classical methods, our approach significantly reduces the error rate in neuronal topology reconstruction. The code and dataset have been released in  \href{https://github.com/FullIsCool/embed_net}{https://github.com/FullIsCool/embed\_net}.
\end{abstract}

\section{Introduction}

Automatic segmentation and three-dimensional reconstruction of individual neurons are key to deciphering the brain’s microscale connectome and functional networks. The morphology of a single neuron’s dendrites and axon, together with the spatial distribution of its synapses, determines the neuron’s input–output patterns. High-precision morphological reconstructions enable the assembly of detailed brain connectomes, thereby revealing the topological organization of sensory, motor, and cognitive circuits \cite{Lichtman2008,Seung2012}.

Fluorescence Micro‐Optical Sectioning Tomography (fMOST) \cite{Wang2021} combines precision mechanical sectioning with high-sensitivity fluorescence imaging to achieve whole-brain, single-cell resolution ($0.35\times0.35\times1.0 \mu m$/voxel). On platforms such as Wuhan OE-Bio BioMap5000, fMOST is highly automated, yet a single mouse brain can generate on the order of 16TB of image data—far exceeding manual tracing capabilities. Even with semi-automated tools \cite{Gapr2024Gou}, annotation throughput lags behind data acquisition, making labeling the primary bottleneck for large-scale quantitative analysis.

\begin{figure}
    \centering
    \includegraphics[width=0.8\linewidth]{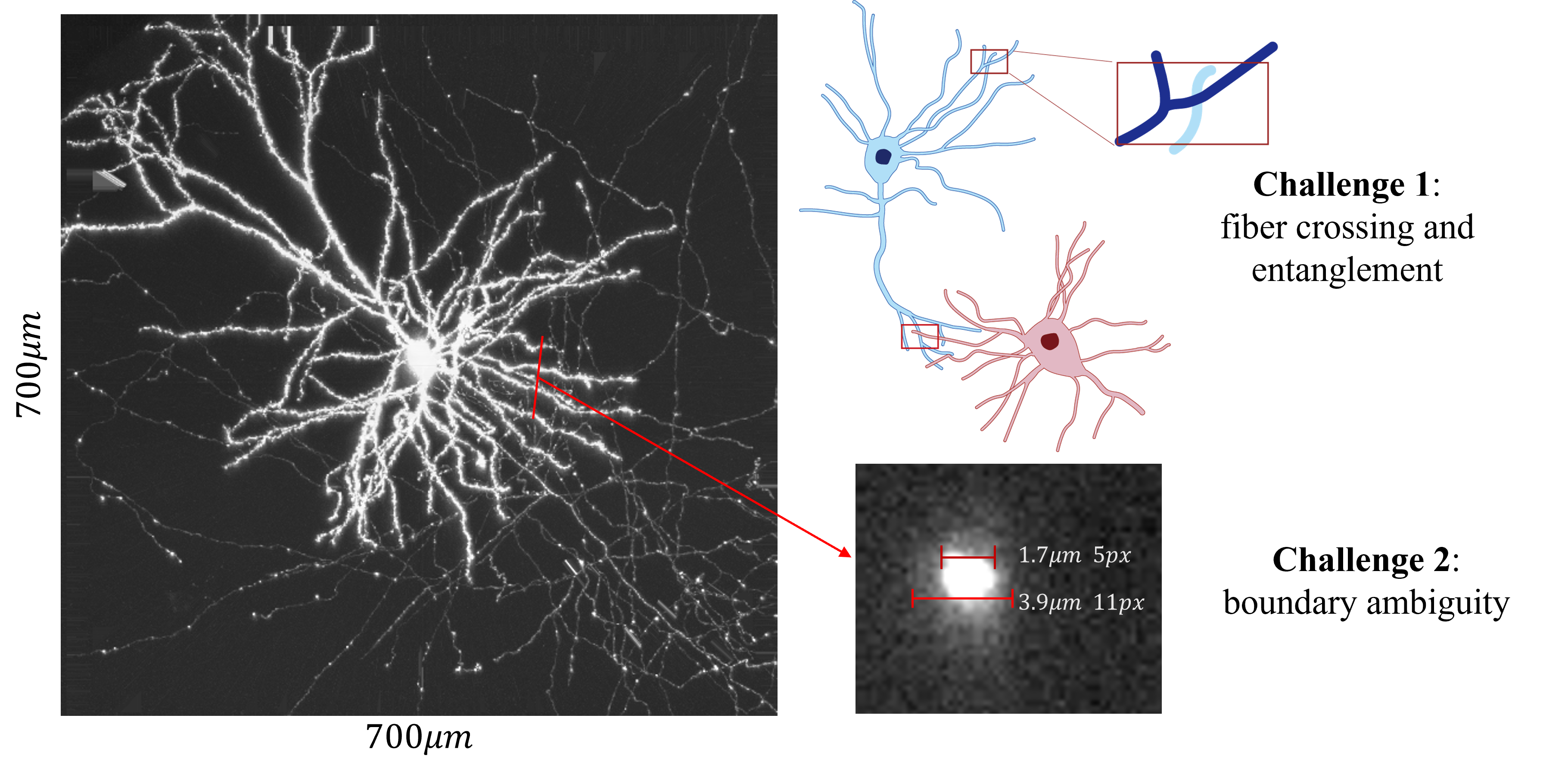}
    \caption{Challenges}
    \label{fig:challenges}
\end{figure}

Recently, deep learning–based methods for automated neuron reconstruction have emerged \cite{Zhou2018DeepNeuron,Callara2020,Li2020}. However, fully automatic, high-fidelity reconstructions remain elusive due to two fundamental challenges. Fig. \ref{fig:challenges} illustrates these two challenges First, fiber crossing and self-entanglement: dendrites and axons often wind around themselves or intersect with processes from other neurons over long distances, a scenario distinct from standard 3D instance-segmentation occlusions. Overlapping branches in the image volume frequently lead two-class semantic segmentation networks to merge unrelated fibers, producing fused boundaries and topological errors. Second, boundary ambiguity and low signal-to-noise ratio: under fMOST, neurite diameters span only a few pixels and are affected by optical diffraction, uneven fluorescence labeling, and background noise, yielding gray-scale transition zones several pixels wide. Such blurred edges impede accurate segmentation and often induce breakpoints or spurious connections during skeleton extraction.

These limitations restrict the applicability of existing approaches to whole-brain datasets and prevent reconstructions from meeting the dual demands of accuracy and completeness in neuroanatomical and connectomic studies.

To address these challenges, we propose an end-to-end reconstruction framework based on pixel-level embedding. Our network outputs a high-dimensional feature vector for each voxel, so that distances in embedding space naturally separate different fiber segments and avoid the branch-merging pitfalls of binary segmentation. A subsequent graph-based post-processing module further refines connections at crossing and discontinuity sites. Recognizing the critical importance of connectivity in neuronal morphology, we also introduce a multi-scale connectivity evaluation metric that, unlike a single Betti-number measure, quantifies connection errors at various scales. Experiments using this metric demonstrate that our method substantially reduces connectivity errors.

Our main contributions are as follows:
\begin{enumerate}
  \item We reformulate neuron reconstruction as a pixel embedding problem and design a discriminative embedding loss tailored to fiber separation.
  \item We develop an end-to-end reconstruction pipeline that converts raw fMOST volumes directly into SWC-format neuron trees.
  \item We propose a graph-topology-driven connectivity metric that quantifies connection errors at multiple scales, enhancing the biological credibility of reconstructed neurons.
\end{enumerate}

\section{Related Work}
\label{sec:related work}

\noindent\textbf{3D Neuron Reconstruction Methods.}
Since the release of V3D (now Vaa3D) by Peng et al. in 2010\cite{V3D2010Peng}, this platform has become a leading tool for large-scale 3D bioimage visualization and analysis. Vaa3D supports smooth, real‐time browsing of terabyte‐scale, multi‐resolution datasets, fully meeting the demands of whole‐brain, high‐resolution visualization. Building on Vaa3D, the V3D-Neuron plugin has introduced various automatic tracing algorithms. For example, the APP2 (All-Path Pruning) algorithm employs a gray‐weighted distance transform combined with front propagation to assemble tree‐like topologies. The UltraTracer method, grounded in maximum‐likelihood estimation, completes reconstruction through statistical modeling\cite{Peng2017}. In 2019, Li et al.\cite{gcut2019Li} proposed the G-Cut algorithm, which leverages quantified neurite growth orientation features and a graph‐search strategy to provide a globally optimal segmentation of overlapping neuronal branches.

Manual, expert‐guided annotation remains the gold standard for accurate neuron morphology reconstruction. Engineering optimizations have significantly improved annotation efficiency. In 2017, Gou et al.\cite{FNT2017Gou} developed the Fast Neuron Tracer (FNT), which integrates the entire tracing process into a cohesive software suite aligned with human annotation logic, resulting in the largest corpus of 6,357 mouse prefrontal cortex projection neurons to date\cite{6357PFC2022Gao}. More recently, the Gapr software\cite{Gapr2024Gou} introduced further enhancements, including multi‐user, multi‐device annotation support and U-Net–based AI assistance.\\

\noindent\textbf{Deep Learning–Based Methods.}
Ronneberger et al.\cite{unet2015Ronneberger} first introduced the U-Net architecture to address biomedical segmentation tasks characterized by small foregrounds and stringent accuracy requirements. Çiçek et al.\cite{3dunet2016Cicek} extended U-Net to volumetric data (3D U-Net). Subsequent works have proposed variants such as U-Net++\cite{UnetPP2018Zhou}, Attention U-Net\cite{attention2018oktay}, and ResUNet\cite{ResUNet2020Foivos}, enhancing feature extraction through revised backbones, refined skip connections, or attention modules.

Given the tubular geometry of vessels and neurites, specialized convolutional operations have been explored: Yu et al.\cite{Dilated2017Yu} employed dilated convolutions to enlarge the receptive field; Dai et al.\cite{Deformable2017Dai} introduced deformable convolutions to adaptively capture geometric deformations; Qi et al.\cite{Dynamicsnake2023Qi} designed dynamic snake convolutions for smooth curve modeling. These modules were integrated into U-shaped backbones to create networks such as D-UNet\cite{DUNet2019Jin}, DeUNet\cite{DeUNet2022Dong}, and DCU-net\cite{DCU-net2022Yang}, demonstrating marked improvements on public datasets of retinal and myocardial vasculature.

Other studies have focused on loss functions tailored for tubular structure segmentation. Shit et al.\cite{clDICEShit2021} proposed the clDICE similarity measure to strengthen tubular recognition. Wang et al.\cite{DDT2020Wang} introduced Deep Distance Transform (DDT), a geometry‐aware segmentation framework. While these methods emphasize continuity constraints, inaccuracies and offsets in the extracted skeleton can undermine their effectiveness. Leveraging persistent homology, King et al.\cite{topoloss2022King} formulated a topological loss function that enforces topological similarity during training to restore continuity in fragmented regions.

\section{Methodology}
\label{sec:Methodology}

\subsection{Pixel embedding U-Net}
Based on the 3D U-Net architecture of Ronneberger et al. \cite{3dunet2016Cicek}, we modify the final layer to produce an $n$-dimensional embedding for each voxel. Specifically, we replace the original single-channel $1\times1\times1$ convolution with a $1\times1\times1$ convolution having $n$ output channels. The remainder of the encoder–decoder backbone retains its four-level symmetric U-shape, thereby preserving 3D U-Net’s strength in foreground/background binary segmentation.

Our model adopts a dual-branch output:

Segmentation branch: maintains the original foreground/background classification task, trained with a binary cross-entropy loss to enforce accurate encoding and decoding.  

Embedding branch: outputs an $n$-dimensional feature vector for each voxel and employs a metric-learning loss to encourage tight clustering of vectors within the same fiber segment and clear separation between different segments.

\begin{figure}
    \centering
    \includegraphics[width=1.0\linewidth]{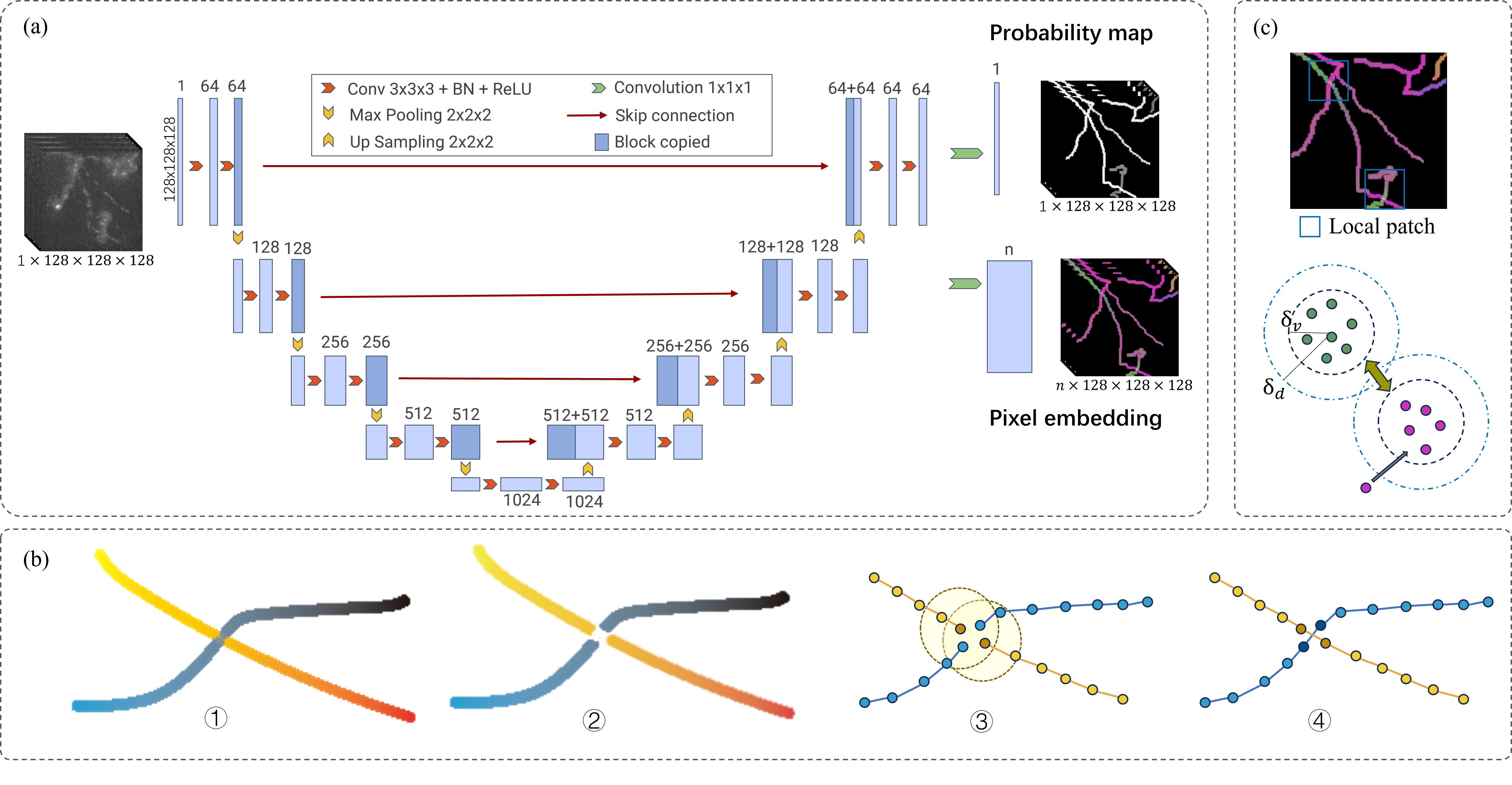}
    \caption{(a)Schematic diagram of the model structure using pixel embedding. (b)post-processing for generating SWC. (c)Embedding vector clustering within a local field of vision }
    \label{fig:model}
\end{figure}

As illustrated in Fig.\ref{fig:model}(a), the network implements a differentiable mapping
\begin{equation}
\begin{aligned}
  &f_{\theta}\colon \mathbb{R}^{H\times W\times D}
    \longrightarrow \mathbb{R}^{n\times H\times W\times D},\\
  &\mathbf{X} = f_{\theta}(\mathbf{V}),\quad 
    \mathbf{X} \in \mathbb{R}^{n\times H\times W\times D}, \quad
  \mathbf{x}_i \in \mathbb{R}^n
    \quad\text{for each voxel }i.
\end{aligned}
\end{equation}
where $\mathbf{v}_i$ denotes the $i$-th input voxel and $\mathbf{x}_i$ is its corresponding $n$-dimensional embedding. The complete set of outputs $\{\mathbf{x}_i\}$ thus provides highly discriminative representations for subsequent neuron segmentation and connectivity analysis.

\subsection{Discriminative Loss Based on Local Vector Clustering}

To effectively separate interleaved and overlapping neurite segments in the high-dimensional embedding space, we adapt the discriminative loss of De Brabandere et al.\cite{DeBrabandere2017} and incorporate the long-range continuity and local occlusion characteristics of neuronal fibers. The key idea is: first, Use prior ground-truth labels to locate small patches where overlaps occur. Then, Within each patch, treat each micro-segment as an independent instance and impose clustering constraints on its embedding vectors.

Our local discriminative loss comprises four terms:
\begin{itemize}
  \item[(i)] Variance term $L_{\mathrm{var}}$ (intra‐cluster compactness)
  \item[(ii)] Distance term $L_{\mathrm{dist}}$ (inter‐cluster separation)
  \item[(iii)] Continuity term $L_{\mathrm{con}}$ (smoothness along fibers)
  \item[(iv)] Regularization term $L_{\mathrm{reg}}$ (embedding norm penalty)
\end{itemize}

Let $\delta_v$ and $\delta_d$ be the intra‐cluster and inter‐cluster margins, respectively. We denote by $C_k$ the number of instances in local patch $k$, by $S_c$ the set of voxels belonging to instance $c$, and by $\mu_c$ the centroid of those voxels' embedding vectors. The variance term measures the distance between each voxel’s embedding and its instance centroid, enforcing $\|x_i - \mu_c\| \le \delta_v$
to ensure adequate intra-cluster compactness. The distance term ensures that centroids of different instances are well separated in the embedding space by requiring$\|\mu_c - \mu_{c'}\| \ge 2\,\delta_d$. As is explained in Fig.\ref{fig:model}(c).

\begin{equation}\label{eq:Lvar}
  L_{\mathrm{var}}^{(k)}
  = \frac{1}{C_k}
    \sum_{c=1}^{C_k}
    \frac{1}{|S_c|}
    \sum_{i\in S_c}
    \bigl[\|x_i - \mu_c\| - \delta_v \bigr]_+^2,
\end{equation}

\begin{equation}\label{eq:Ldist}
  L_{\mathrm{dist}}^{(k)}
  = \frac{1}{C_k(C_k-1)}
    \sum_{\substack{c_A,c_B=1\\c_A\neq c_B}}^{C_k}
    \bigl[\,2\delta_d - \|\mu_{c_A}-\mu_{c_B}\|\,\bigr]_+^2,
\end{equation}
where $\|\cdot\|$ denotes the Euclidean norm and $[x]_+ = \max(0,x)$. The superscript $k$ indexes patches.

To enforce smooth variation along each neurite, let $A(x_i)$ be the set of embedding vectors in the 26-neighborhood of $x_i$. The continuity term is

\begin{equation}\label{eq:Lcont}
  L_{\mathrm{con}}
  = \frac{1}{2\,|\{x_i\}|}
    \sum_{x_i}
    \frac{1}{|A(x_i)|}
    \sum_{x_i' \in A(x_i)}
    \bigl[\|x_i' - x_i\| - \delta_v\bigr]_+^2.
\end{equation}

To prevent feature explosion, we add a regularization term:

\begin{equation}\label{eq:Lreg}
  L_{\mathrm{reg}}
  = \frac{1}{|\{x_i\}|}
    \sum_{x_i}\|x_i\|^2.
\end{equation}

During training, we also include the binary cross‐entropy loss $L_{\mathrm{bce}}$ from the segmentation head. The total loss is

\begin{equation}\label{eq:Ltotal}
  L_{\mathrm{total}}
  = \frac{1}{N_{\mathrm{patch}}}
    \sum_{k=1}^{N_{\mathrm{patch}}}
      \bigl(\alpha\,L_{\mathrm{var}}^{(k)} + \beta\,L_{\mathrm{dist}}^{(k)}\bigr)
    + \gamma\,L_{\mathrm{con}}
    + \eta\,L_{\mathrm{reg}}
    + \xi\,L_{\mathrm{bce}},
\end{equation}
where $\alpha,\beta,\gamma,\eta,\xi$ are weighting hyperparameters and $N_{\mathrm{patch}}$ is the total number of local patches. This design ensures that crossing fibers are locally well separated while maintaining global smoothness and compactness of each neurite’s embedding.

\subsection{Post-processing methods for establishing the neural fiber skeleton}

In neuro‐morphological studies, the skeleton representation encapsulates the essential topological information of neuronal processes. Specifically, a series of nodes are placed along each neurite, and their parent–child relationships define a tree structure. The SWC (Standardized Wireframe Container) format \cite{Ascoli2007} is widely adopted for quantitative description of 3D neuron morphology. Each SWC record comprises seven fields: node index, type, three‐dimensional coordinates, radius, and parent node index, thereby fully encoding the spatial connectivity of the neuron. Compared to detailed geometric meshes, the skeleton more directly reflects topological features such as branching patterns and hierarchy, making it especially valuable for morphometric analyses.

To reconstruct neurons from the learned embeddings, we first apply our retrained segmentation network to obtain a binary foreground mask and compute an $n$‐dimensional embedding vector for every voxel. Next, we impose an embedding‐distance threshold $\epsilon$ on each pair of adjacent foreground voxels: if the Euclidean distance between their embeddings exceeds $\epsilon$, the link is deemed a cross‐fiber break. This criterion partitions the foreground mask into a collection of micro‐segments. We then perform 3D parallel thinning \cite{thinning1998Kalman} on these segments to extract a one‐voxel‐wide skeleton. 

To restore continuity across true neurite crossings and small gaps, we examine each skeleton node’s neighborhood: any two adjacent skeleton voxels whose embedding distance falls below $\epsilon$ are reconnected. Finally, a breadth‐first search (BFS) on the resulting connectivity graph yields a complete tree structure. Fig.\ref{fig:model}(c) illustrates the entire pipeline from embedding field to final skeleton, including the “jump‐connection” strategy at crossing points.
 
\section{Experiments}
\subsection{Connectivity Evaluation Algorithm}
\begin{figure}
    \centering
    \includegraphics[width=0.8\linewidth]{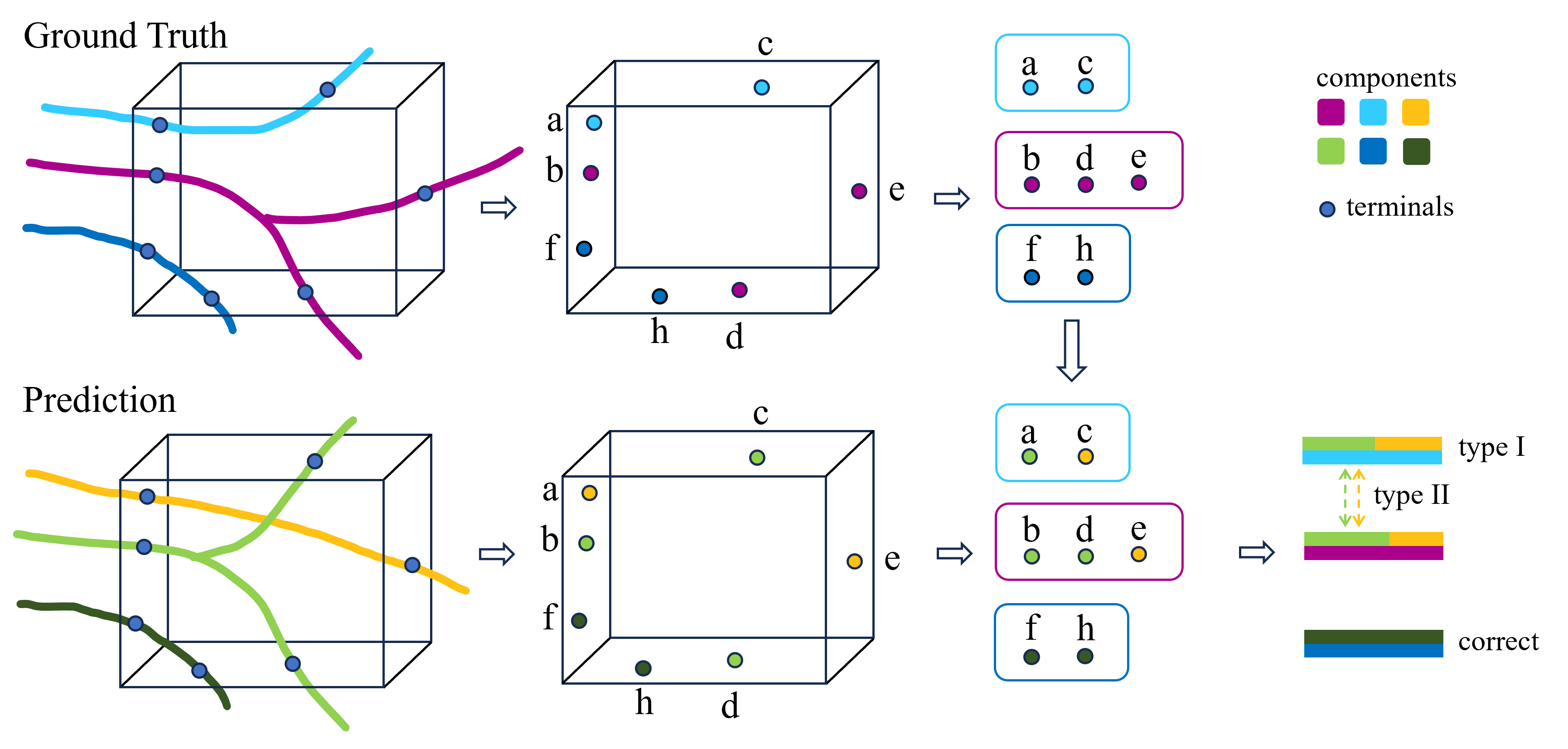}
    \caption{Schematic diagram of connectivity evaluation algorithm}
    \label{fig:evaluation}
\end{figure}

In neuronal connectomics, our focus lies not on microscopic branch details but on the projection directions and global connectivity of neural fibers. As illustrated in Fig.\ref{fig:evaluation}, two reconstructions may share the same Betti number yet exhibit markedly different connection patterns—differences that a single scalar topological metric cannot capture.

To address this, we propose a multiscale connectivity evaluation pipeline:

1. \textbf{End‐Point Extraction.}  
   On both the ground‐truth and predicted SWC trees, identify all nodes located at volume boundaries or branch termini. These constitute the terminal node sets.

2. \textbf{Ground‐Truth Component Partition.}  
   Using the connectivity graph of the ground‐truth SWC tree, partition its terminal nodes into connected subsets. Each subset represents the projection of a single neurite at the boundary. Similarly, partition the predicted terminals according to the predicted SWC connectivity.

3. \textbf{Terminal Pairing.}  
   Establish a one‐to‐one correspondence between predicted and ground‐truth terminals via nearest‐neighbor matching in Euclidean space.

4. \textbf{Error Classification.}  
   – \emph{Type I (Disconnection).} If terminals from a single ground‐truth component are paired with terminals in two or more distinct predicted components, this indicates a false break—i.e., a true connection has been erroneously severed.  
   – \emph{Type II (Overconnection).} If a single predicted component contains terminals paired to two or more different ground‐truth components, this indicates a spurious merge—i.e., fibers that should remain separate are incorrectly connected.
   components are represented by different colors in Fig.\ref{fig:evaluation}.
   
By tallying these two error types, we achieve a fine‐grained, scale‐aware quantification of connectivity fidelity, overcoming the limitations of traditional single‐value topological metrics.

\subsection{Experimental Results}

\noindent\textbf{Datasets.}\\
We constructed an in‐house fMOST neuronal imaging dataset. Ai166 reporter‐line mice were crossed with a layer‐specific Cre driver to sparsely label pyramidal neurons in layers I–V of the prefrontal cortex \cite{Veldman2020}. Whole‐brain imaging was then performed on the BioMap5000 fMOST system (Wuhan OE-Bio ) at $0.35\times0.35\times1.0\ \mu$m voxel resolution \cite{Wang2021}. From three mice, we randomly extracted 180 subvolumes exhibiting strong fluorescence, capturing somata, dendrites, and axons. Using the Fast Neuron Tracer (FNT) software \cite{FNT2017Gou}, expert annotators manually placed skeleton nodes within each subvolume. Each region was then partitioned into $128^3$‐voxel blocks and augmented by random rotations and mirror flips, yielding a dataset of 33,000 fully annotated $128^3$ samples for training and evaluation.\\

\noindent\textbf{Baseline Models.}\\
To assess the topological fidelity afforded by our embedding network, we compared against three standard U‐Net variants: the vanilla 3D U‐Net, U‐Net++ \cite{UnetPP2018Zhou}, and Attention U‐Net \cite{attention2018oktay}. All networks shared an identical channel configuration to ensure fair comparison. This study tests the hypothesis that improvements in pure binary segmentation accuracy alone cannot deliver comparable gains in topological reconstruction without embedding‐based representations.\\

\noindent\textbf{Implementation Details.}\\
We monitor performance along three axes:  
(1) Segmentation quality—Accuracy, Precision, Recall, and F1 score;  
(2) Overlap metrics—Dice coefficient and Intersection over Union (IoU);  
(3) Topological connectivity—our proposed skeleton‐connectivity evaluation.  
The dataset was split 80\%/20\% into training and validation sets. All models were trained with the Adam optimizer, an initial learning rate of $1\times10^{-4}$, and identical remaining hyperparameters.\\

\noindent\textbf{Results.}\\
Table~\ref{tab:comparison} summarizes segmentation and overlap metrics for the four models. Since our embedding network builds upon the same 3D U‐Net backbone for foreground extraction, its segmentation metrics match those of the vanilla 3D U‐Net. U‐Net++ underperforms on our dataset, while Attention U‐Net achieves improvements of 5.2\% in Dice and 3.4\% in IoU. However, because the ground‐truth masks are derived from annotated SWC skeletons, even the baseline U‐Net already attains near‐ideal segmentation—hence further gains offer limited benefits for downstream skeleton extraction.

Table~\ref{tab:connectivity} reports the connectivity error counts. For Type I (disconnection) errors, all well‐segmented models register counts near zero, whereas U‐Net++ hampered by segmentation breaks exhibits a higher Type I rate. For Type II (overconnection) errors, our embedding network cuts the error count by nearly 50\% compared to the best baseline. Overall, with almost no increase increasing computational complexity or parameter count, our method substantially reduces both disconnection and overconnection errors.
\begin{figure}
    \centering
    \includegraphics[width=1.0\linewidth]{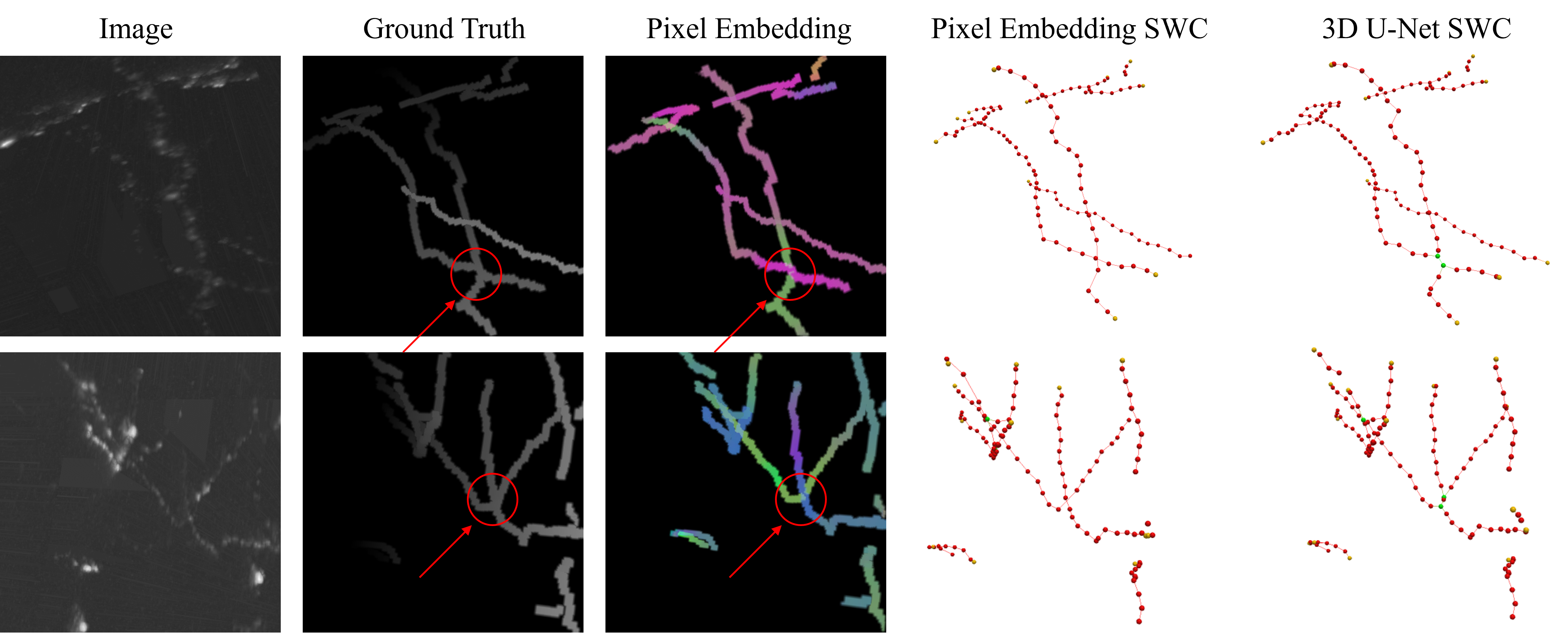}
    \caption{Results demonstration. First column: raw input image (the actual input is a 3D volume; shown here as a 2D projection). Second column: ground truth — to visualize the 3D structure, different depths are rendered in varying shades of gray. Third column: predicted vector field, with high-dimensional embeddings mapped into color space. Fourth column: SWC node connectivity predicted by our method. Fifth column: SWC connectivity predicted by the U-Net baseline; mutually connected branch points are highlighted in green.}
    \label{fig:results}
\end{figure}

\begin{table}[htbp]
  \centering
  \caption{Comparison of Semantic Segmentation Performance}
  \label{tab:comparison}
  \begin{tabular}{lcccc!{\vrule width 0.3pt}cc}
    \toprule
                   & Accuracy & Precision & Recall  & F1-score & Dice    & IoU    \\
    \midrule
    Ours          & 99.93\%  & 85.74\%   & 95.35\% & 85.53\%  & 0.8553  & 0.8170  \\
    3D U-Net       & 99.93\%  & 85.74\%   & 95.35\% & 85.53\%  & 0.8553  & 0.8170  \\
    U-Net++        & 99.94\%  & 82.89\%   & 74.25\% & 73.76\%  & 0.7376  & 0.6208  \\
    Att U-Net      & 99.98\%  & 91.22\%   & 92.84\% & 90.76\%  & 0.9076  & 0.8515  \\

    \bottomrule
  \end{tabular}
\end{table}

\begin{table}[htbp]
  \centering
  \caption{Comparison of Models'c Connection Performance}
  \label{tab:connectivity}
  \begin{tabular}{lccccc}
    \toprule
                   & correct    & typeI     & typeII  & typeI/correct       & typeII/correct   \\
    \midrule
    Ours          & \bfseries15.2       & 0.03      & 1.62    & \dashuline{$2.0\times10^{-3}$} & \bfseries0.106  \\
    3D U-Net       & 13.7       & 0.006     & 3.18    & \dashuline{$4.4\times10^{-4}$}  & 0.232  \\
    U-Net++        & 13.6       & 3.3       & 1.97    & 0.24                & 0.145  \\
    Att U-Net      & 14.3       & 0.006     & 2.56    & \dashuline{$4.2\times10^{-4}$}   & 0.179     \\

    \bottomrule
  \end{tabular}
\end{table}

\begin{table}[htbp]
  \centering
  \caption{Model parameter quantity and inference time (running on NVIDIA A800 80 GB Tensor Core GPU)}
  \label{tab:models}
  \begin{tabular}{lcc}
    \toprule
                   & Params  & Inference time(s)  \\
    \midrule
    Ours           & 90.4M       & 0.054        \\
    3D U-Net       & 90.3M       & 0.039       \\
    U-Net++        & 104.8M      & 0.152   \\
    Att U-Net      & 91.4M       & 0.068      \\

    \bottomrule
  \end{tabular}
\end{table}

\section{Conclusion}

This study systematically optimizes connectivity errors in automatic neuron reconstruction under fMOST imaging. We first introduce a deep network that outputs pixel‐level embedding vectors and design a discriminative loss to ensure that, even in regions of overlap and occlusion, different neuronal fibers are effectively separated in the high‐dimensional feature space. Building on this model, we develop an end‐to‐end reconstruction pipeline that converts raw volumetric data directly into SWC‐formatted skeleton trees. To overcome the limitations of traditional metrics in assessing structural fidelity, we propose a novel topological evaluation metric that quantifies segmentation and reconstruction quality at multiple scales and dimensions. Experimental results demonstrate that our method significantly reduces connectivity errors. Nevertheless, in exceedingly complex fiber‐entanglement scenarios, current U‐Net‐based architectures remain insufficient, underscoring the need for more innovative algorithms to further enhance reconstruction robustness.

\newpage
\bibliographystyle{unsrt}
\bibliography{refs}
\end{CJK}
\end{document}